# Active polarization control for quantum communication in long-distance optical fibers with shared telecom traffic

**G. B. Xavier,[1,*] G. Vilela de Faria,[1] T. Ferreira da Silva,[1,2] G. P. Temporão[1] and J. P. von der Weid[1]**

[1]*Center for Telecommunication Studies, Pontifical Catholic University of Rio de Janeiro, R. Marquês de São Vicente 225 – Rio de Janeiro - Brazil*

[2]*National Institute of Metrology, Standardization and Industrial Quality, Av. Nossa Sra. das Graças, 50 - Duque de Caxias - Brazil*

[*]*Corresponding author: guix@opto.cetuc.puc-rio.br*

We experimentally demonstrate the compatibility of wavelength multiplexed active polarization stabilization for quantum communication in an optical fiber carrying telecom traffic. One of the feedback control channels contains a 9.953 Gb/s data stream generated from a BER meter. We verify the ability to transmit single-photons in the two opposite directions of a 23 km optical fiber spool, while maintaining their state of polarization stable and a classical BER in the feedback channel error-free, during 6 hours of continuous operation.

## 1. Introduction

Quantum communication [1] has changed the way we look at the simple act of information transfer between two remote parties. From the first experiment of quantum key distribution (QKD) [2], to experimental demonstrations of quantum repeaters [3], the field of experimental quantum communication is expanding at an increasing pace. Quantum communication typically employs single-photons with one of their degrees of freedom encoding the information, usually refereed to as qubits [4]. In order to carry the quantum signals from the transmitter to the receiver (Alice and Bob respectively), a suitable transmission channel is needed. Most quantum communication experiments to date have used either optical fibers [5-8] or free-space links [9-11] as the channel.

As quantum communication moves out of the lab and into practical applications, there has been a recent interest in performing it in a classical fiber-optical network environment [12-13]. Some of these experiments have also performed quantum key distribution (QKD) in an optical fiber shared with strong classical signals using wavelength multiplexing, improving the integration between classical and quantum communications [14-15]. Nevertheless careful filtering and limited classical launch powers must be employed to avoid degradation of the qubits due to scattering and non-linear effects generated by the classical signals inside the optical fiber [15-16].

Research on the integration of quantum communication systems into classical optical telecom network environments, is of great interest, as it may allow the wide deployment of this new technology in the future [15]. Optical fibers already make excellent channels for the propagation of single-photons. The same benefits optical fibers offer to classical communications, such as extreme low loss and high bandwidth potential, are also available for quantum communications. One extra advantage, which is usually not a problem for classical telecom, is that single-photons have a very low probability of undergoing decoherence (coupling of the quantum state to the environment, usually an irreversible process) while propagating in a fiber [4]. Since both classical and quantum systems share the same fiber, they are typically wavelength multiplexed, taking advantage of the existing ITU-T dense wavelength division multiplexing (DWDM) wavelength grid for easier compatibility with commercial systems [15].

We have recently demonstrated an active polarization control system that is able to undo any changes to the state of polarization (SOP) caused by the residual birefringence changes in an optical fiber link [17]. Our system uses two classical continuous-wave (CW) feedback signals, wavelength multiplexed with the single-photons, and has been proven effective in a QKD experiment using polarization encoded qubits [18]. We demonstrate in this letter, that it is possible to replace the CW classical channels with high-speed telecom-grade data signals showing the compatibility of our control system with classical data transmission, while still simultaneously transmitting single-photons with active polarization stabilization. We also show that it is possible to send the single-photons co or counter-propagating with the classical channels, allowing for greater flexibility of the application of the active control in an optical fiber with telecom traffic present.

## 2. Experimental setup

The entire experimental setup employs standard off-the-shelf fiber-pigtailed telecom components, and is shown in Fig. 1. We have employed the usual nomenclature of quantum communication in our experiment, naming Alice and Bob the communicating parties. The classical telecom information propagates in the fiber from Alice to Bob, while the single-photons, stabilized against polarization drifts, are transmitted in both directions. Our polarization control system requires that two classical channels are used, with their SOPs aligned non-orthogonally [17], wavelength multiplexed with the quantum signal. At the other fiber end the intensities of each control channel are individually measured after linear polarizers, oriented with manual polarization controllers, such that they individually match the corresponding channel input states. These two intensities are used as the feedback signal to the control system. The channels are located 0.8 nm apart in the 1550 nm window, corresponding to the ITU-T 100 GHz grid spacing. The quantum channel is placed at $\lambda_Q$ = 1546.12 nm, with the control channels at $\lambda_1$ = 1545.32 nm and $\lambda_2$ = 1546.92 nm.

The laser at $\lambda_1$ is a standard telecom (Distributed Feedback) DFB laser diode operating in (continuous wave) CW mode. The one at $\lambda_2$ is the output of a DFB laser from a BER (bit error rate) test meter generating a $2^{31}$-1 pseudo-random bit sequence operating at 9.953 Gb/s, simulating real telecom traffic. Both channels are generated at Alice and pass through manual polarization controllers to adjust their individual SOPs, and band-rejection filters (BRF, two fiber Bragg gratings - FBGs, in series, yielding > 60 dB extinction ratio between adjacent channels, with an optical isolator in between them to avoid the formation of a cavity). These BRFs are centered at $\lambda_Q$ = 1546.12 nm, and are used to remove photons generated from the amplified spontaneous emission (ASE) noise generated by the laser, which would fall in-band with the quantum channel.

Finally, the two signals are combined by a DWDM filter before the optical link together with $\lambda_Q$ (isolation between adjacent channels > 40 dB), bringing the total isolation between the classical and quantum channels to > 100 dB. In order to minimize noise from Raman spontaneous scattering, we use launch powers for each channel of -19.8 dBm after the DWDM. The launch powers are afterwards compensated at Bob with erbium doped fiber amplifiers (EDFAs), together with band-pass filters to diminish the noise generated from the EDFAs. This is the minimum power level that guaranteed a classical BER free of errors with our experimental setup. Note that this is far below the usual value for input powers for telecom systems (0 dBm), but without doing so Raman scattering induced noise increases by a factor of ~ 95 since it increases linearly with input power [16]. The DWDMs used by both Alice and Bob are identical, have over 40 dB isolation between adjacent channels and ~ 1.6 dB insertion loss each.

The lasers at $\lambda_Q$ located in both Alice and Bob's stations are identical standard telecom DFB lasers, and for the sake of simplicity for this experiment, they operate in CW mode. The hardware to prepare and analyze the pseudo-single photons in both ends of the link is identical. They are called pseudo single-photons because as they come from an attenuated coherent source, they are only approximations to the ideal single-photon state [4]. Following the lasers there is a manual polarization controller to prepare the transmitted SOPs and an optical attenuator, setting the power level to an average of 0.5 photon per detection window of 2.5 ns in the fiber input. Another manual polarization controller together with a polarizing beam splitter (PBS) is used to analyze the incoming quantum signal from the opposite direction. As shown in Fig. 1, optical circulators are used to correctly route the single-photons in both directions. Filtering is done using FBGs located at $\lambda_Q$ (isolation > 35 dB) and the DWDMs. The combined filtering bandwidth for the quantum channel in both directions is 0.4 nm (50 GHz). It was verified that no cross-talk generated from the classical channels took place by performing photon counting measurements in both directions with the fiber spool disconnected. Finally the two output ports of the PBS are connected to commercial single-photon counting modules (SPCMs). They operate in Geiger mode [4] at a gate frequency of 100 kHz, with a quantum efficiency of 15% at 1550 nm and a gate width of 2.5 ns. The measured dark count probability is 3.7 x $10^{-5}$ and 3.2 x $10^{-5}$ per gate for each SPCMs respectively.

Alice and Bob are separated by 23 km of DS spooled fiber with a mean differential group delay of 0.22 ps. The SOPs of the three channels are actively compensated by a $LiNbO_3$ polarization controller (PC) located at Bob. An incandescent light bulb is used to randomly heat the fiber to simulate temperature changes. Inside Bob's station, the classical channels are split and pass through linear polarizers $P_1$ and $P_2$ oriented at 45º from each other using manual polarization controllers [17]. $\lambda_1$ and $\lambda_2$ are amplified with EDFAs before detections at classical *pin* photodiodes $D_1$ and $D_2$ (300 kHz bandwidth). Note that with the bandwidth of $D_2$, the high-speed telecom data contained in $\lambda_2$ is not at all detected by the control system. $\lambda_2$ is also split by a 50/50 coupler so that it feeds the receiver unit of the BER test meter to monitor the error rate of the 9.953 Gb/s data stream. The two electrical outputs of $D_1$ and $D_2$ are fed back into the control computer to close the feedback loop, allowing us to undo any birefringence rotations the fiber may cause for any input SOP [17, 18].

## 3. Results and discussion

The SOP of the pseudo-single photons at the fiber input at Bob, together with the manual polarization controllers before the PBS at Alice, are adjusted to maximize the counts on one of the SPCMs at Alice, corresponding to the case where the single-photon's SOP sent by Bob matches the measurement basis at Alice [4]. The counts are recorded as a function of time, and the results are plotted in Fig. 2, with an integration time for the detectors of 1 s per measured data point. For the first run, the active control system is switched off and the counts recorded by SPCMs 1 and 2. These have the additional “UC” labels in Fig. 2, indicating they are uncontrolled runs. We observe that the photon count rate at each detector almost immediately begin to drift due to the random birefringence fluctuations. The system is then switched on, the setup realigned and the counts recorded over a similar time period, resulting in the SPCM 1 and 2 curves in the figure. We can clearly observe the stability provided by the control system. The two SPCMs are then connected to the output port of the PBS inside Bob (with the labels now renamed SPCM 3 and 4), the transmitted polarization states are now prepared at Alice, such that once again, the counts are maximized in one detector and minimized in the other at Bob. The counts are then recorded and the results also plotted in Fig. 2 as the curves SPCM 3 and SPCM 4. Once again, the recorded counts remain stable throughout the experimental run when the active polarization control is kept on. The long-term drift mainly observed in the SPCM 1 curve is believed to have been caused by a small change in the SOPs before or after the DWDMs since only the fiber spool is actively controlled. This type of drift can be further reduced by insulating the fiber pigtails in the different components not being controlled from temperature changes in the environment, and also by using an extra feedback to adjust the parameters of the control system when the visibility drops below a certain level. The fluctuations on the recorded curves come from the intrinsic statistical fluctuations of the source / detection process, and as we can observe, when comparing the controlled to the uncontrolled curves, our polarization control system does not add significant noise to the quantum transmission.

The reason why there is roughly a factor two between the count rates between both sets of detectors (1, 2 and 3, 4) is because the attenuation in the two directions is asymmetrical. The average number of photons per detection window entering the fiber is the same, however in the path from Bob to Alice (B-A) the attenuation introduced by the $LiNbO_3$ PC (~ 3dB) is already taken into account in the optical attenuator inside Bob. In the opposite direction (A-B), the attenuation introduced is part of the link and cannot be removed if the same average photon number is to be kept in the fiber input.

It should be noted that our polarization stabilization system is able to compensate for *any* input polarization state as in a true quantum key distribution experiment [18], and as was also shown, it does not add significant noise to the quantum transmission taking place. In fact when the adjustment regarding the input SOPs were made at both Alice and Bob's, no particular input state was chosen, as long as maximum counts on one of the SPCMs are obtained.

One important measure of performance for quantum communication systems is the optical visibility of the photon counts after propagation. It can be expressed as a function of photon-counts as $V = \left|(C_1 - C_2)/(C_1 + C_2)\right|$, where $C_1$ and $C_2$ are the photon-counts per unit time detected at the two SPCMs after the PBSs at both Alice and Bob. The two controlled cases (A-B and B-A directions) exhibit average visibilities of 0.916 ± 0.025 and 0.931 ± 0.016 respectively, with the deviation from perfect visibility stemming from detector dark counts, Raman scattering noise, fluctuations added by the polarization stabilizing system, imperfect manual alignment of the PBSs with the single photons SOPs and limited extinction ratio of the PBSs. Through the use of narrower filters and lower launch power for the classical channels it is possible to reduce the Raman noise [19]. The average visibilities (in the same directions respectively) increase to 0.9522 and 0.9306 if Raman spontaneous noise is lowered to negligible levels, and 0.9677 and 0.9592 if the SPCMs dark counts are subtracted (corresponding to the net visibility). Unsurprisingly, the visibility for the uncontrolled measurement wanders randomly across many possible values, showing once again that quantum communication using polarization states without active control is unfeasible. No noticeable cross-talk occurs in the quantum channel between both directions as the connector reflections and Rayleigh back-scattering are typically at least 30 dB below the propagating quantum signal.

We now plot in Fig. 3 the statistical distribution of the calculated raw visibilities for both controlled directions using the data from Fig. 2. As expected, the Bob - Alice direction has better results due to the lower attenuation.

In order to find out how much farther we can reliably perform polarization encoded quantum communication while simultaneously transmitting classical information, we calculate the visibility as a function of extra fiber distance for both co and counter-propagating cases. Raman noise contribution is calculated by integrating the spontaneously generated photons over the entire length of fiber [19]. The classical BER was also measured using an extra optical attenuator before the receiver simulating increased fiber attenuation. The results are plotted in fig. 4 and were done for two different classical channel input powers: -17.8 dBm and -19.8 dBm (the one used in the experiment). Depending on the input optical power used, the classical BER improves for the same link length, but on the other hand the single-photon visibility decreases due to increased Raman noise.

The visibility for the co-propagating (Alice - Bob) case does not degrade as rapidly as the counter-propagating (Bob - Alice) one since Raman spontaneous noise decreases in the co-propagation direction as the fiber distance increases. In the counter-propagating case this is no longer true since Raman backscattered noise always increases with extra fiber distance [19]. Considering a visibility of 0.8 (corresponding to a QBER of 10%, a practical maximum value for secure key transmission [4]), the maximum transmission distance is ~ 58 km for both directions for -19.8 dBm input power. The counter-propagating visibility is only higher for lower distances because of the asymmetrical attenuation. For the case of -17.8 dBm input power, the maximum supported distance for quantum communication is 53 km in the counter-propagating direction, as shown in Fig. 4. Such a visibility is barely enough for QKD but still high enough to allow Bell inequality violations [20], for example, in a configuration in which one of the parties keeps their photon of an entangled pair, and sends the other one to the other party [12].

From the point of view of the BER results, we have a limit of ~ 3.9 x $10^{-3}$ to the error rate if enhanced forward error correction (EFEC) is used [21]. We observe from Fig. 4 that we are still well within the limit of classical communications using EFEC, and the maximum allowed distance for our setup would be ~ 70 km by extrapolating the error rate curve.

Finally if standard fiber (SMF-28) were to be used in the fiber link together with a dispersion pre-compensator module to allow the classical data transmission to be done, our calculations show that another 10 km could be gained in the quantum transmission distance since less Raman noise would be generated [4]. This would bring the estimated maximum transmission distance for the single-photons to be in line with the results estimated for the classical communication. The polarization control quality is not expected to degrade in these longer distances [17].

## 4. Conclusions

We have shown that active polarization stabilization for quantum communication is compatible with transmission of classical telecom data along the same optical fiber. It is shown that the classical data channel itself could be used to provide feedback to the polarization control system. We have also shown that the transmission of the single-photon states is possible in both directions of the fiber, allowing for a greater flexibility of the integration between the active control and fiber optical telecommunication systems. Limitations of distance were estimated for our current setup performing simultaneous quantum and classical communications. These results are important for applications of active polarization stabilization for quantum communications in optical fibers carrying shared telecom traffic.

The authors acknowledge financial support from the Brazilian agencies CAPES, CNPq and FAPERJ.

## Figure captions

Fig. 1: Experimental setup. The light shaded bounded areas inside both Alice and Bob's stations represent the classical transmission and detection systems. The classical channels propagate from Alice to Bob while the quantum signals propagate simultaneously in both directions. ATT: Optical attenuator; BER Rx and Tx: Bit error rate meter receiver and transmitter modules, respectively; BRF: Band-rejection filter; D1 and D2: Classical *pin* photodetectors; DWDM: Dense wavelength division multiplexing filter; EDFA: Erbium-doped fiber amplifier; FBG: Fiber Bragg grating; ISO: Optical isolator; SPCM: Single-photon counting module; P1 and P2: Linear polarizers; PC: $LiNbO_3$ polarization controller; PBS: Polarizing beam splitter; C: bidirectional 50:50 fiber coupler; FBG: Fiber Bragg-grating filter. Dashed black lines represent electrical connections.

Fig. 2: Experimental photon count results with and without polarization control. SPCM 1 UC and SPCM 2 UC are measurements with the system control turned off.

Fig. 3: Distribution of calculated visibilities obtained from the data in Fig. 3 for Alice - Bob and Bob - Alice directions. Both distributions have the same number of data points.

Fig. 4: Results for the BER of the classical channel and the calculated visibilities for the A-B (Alice - Bob) and B-A (Bob - Alice) directions as function of the link length. H and L correspond to -17.8 and -19.8 dBm classical channel powers respectively. Fiber distance is calculated assuming a mean fiber attenuation of 0.2 dB/km @ 1550 nm.

Figure 1

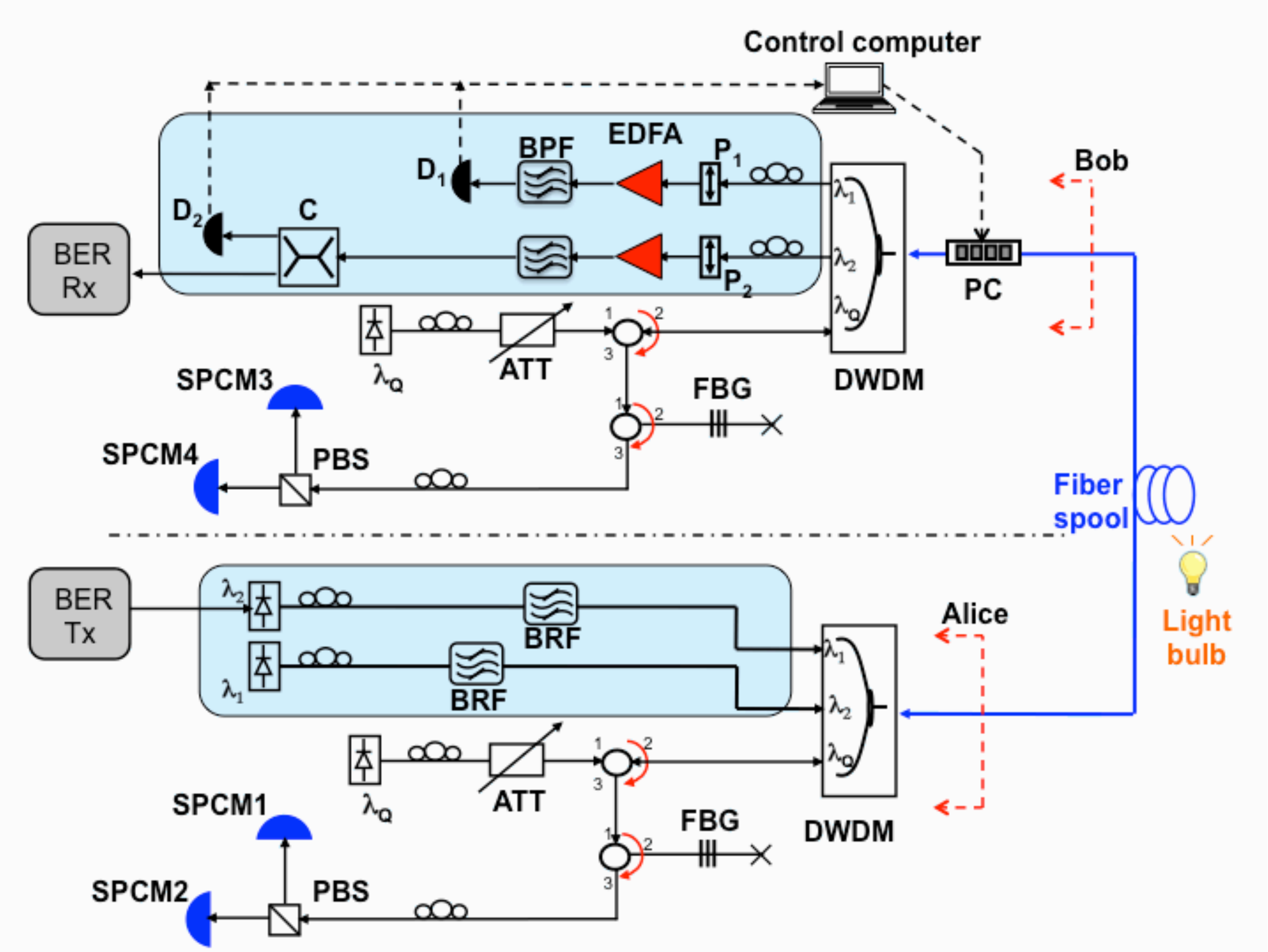

Figure 2

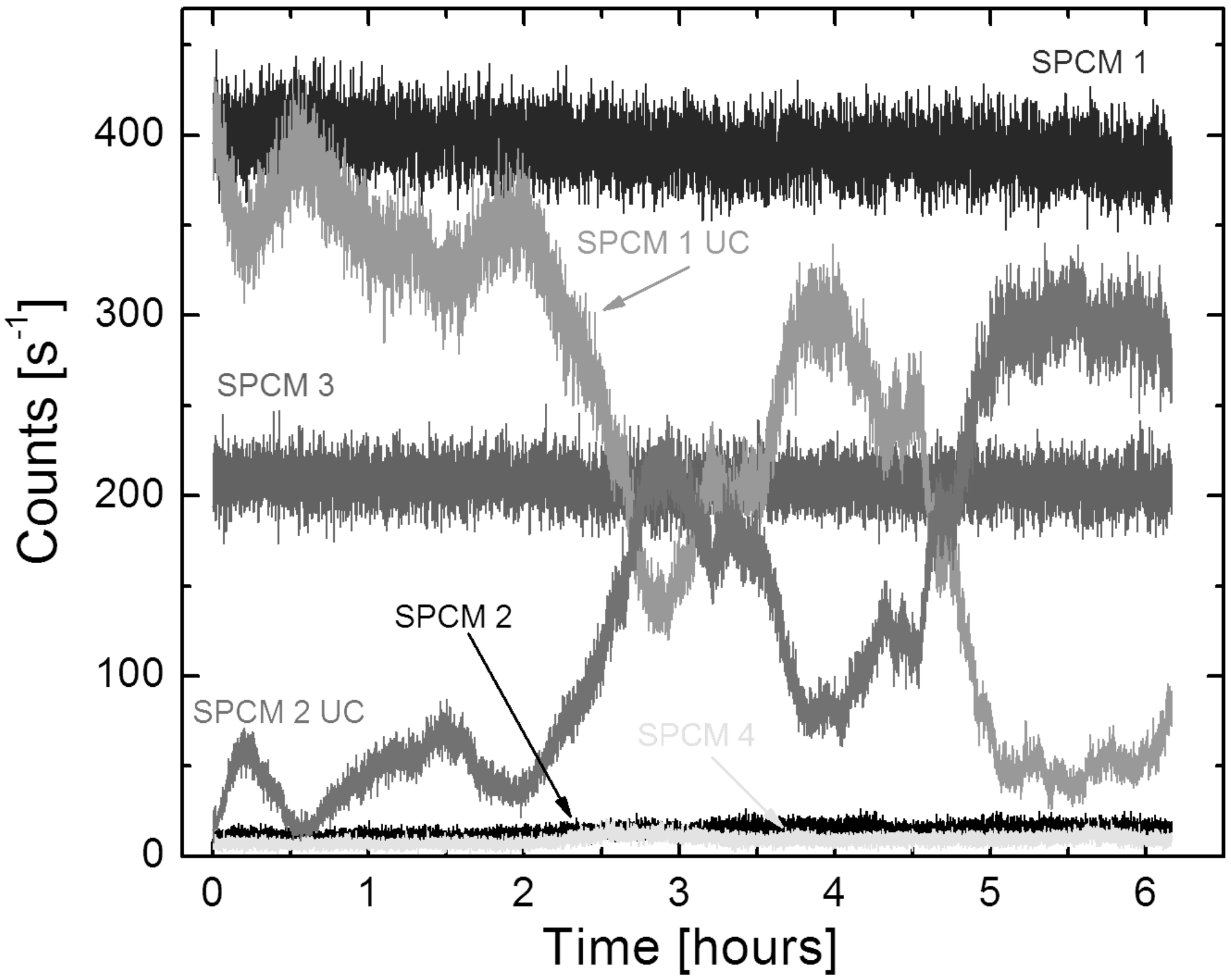

Figure 3

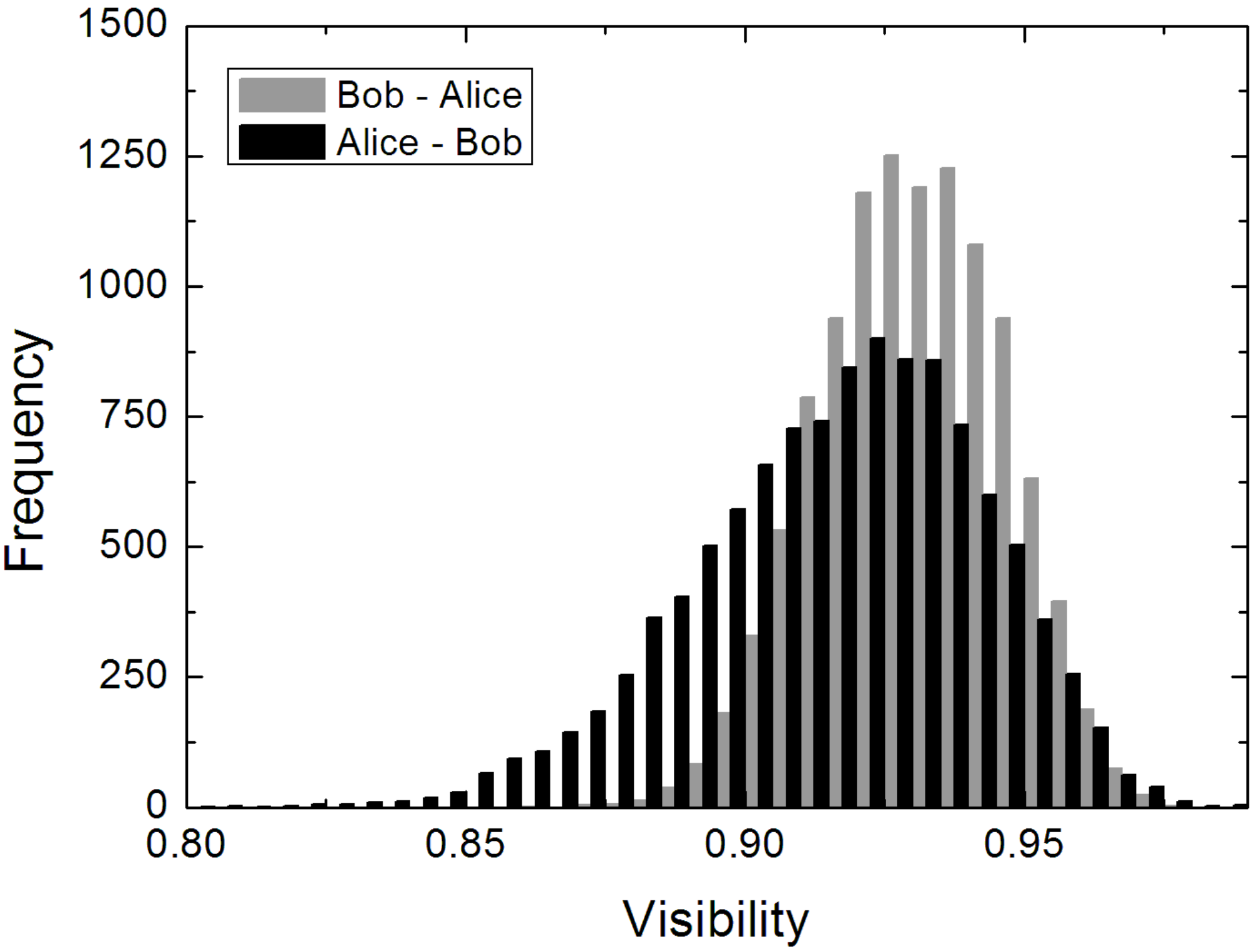

Figure 4

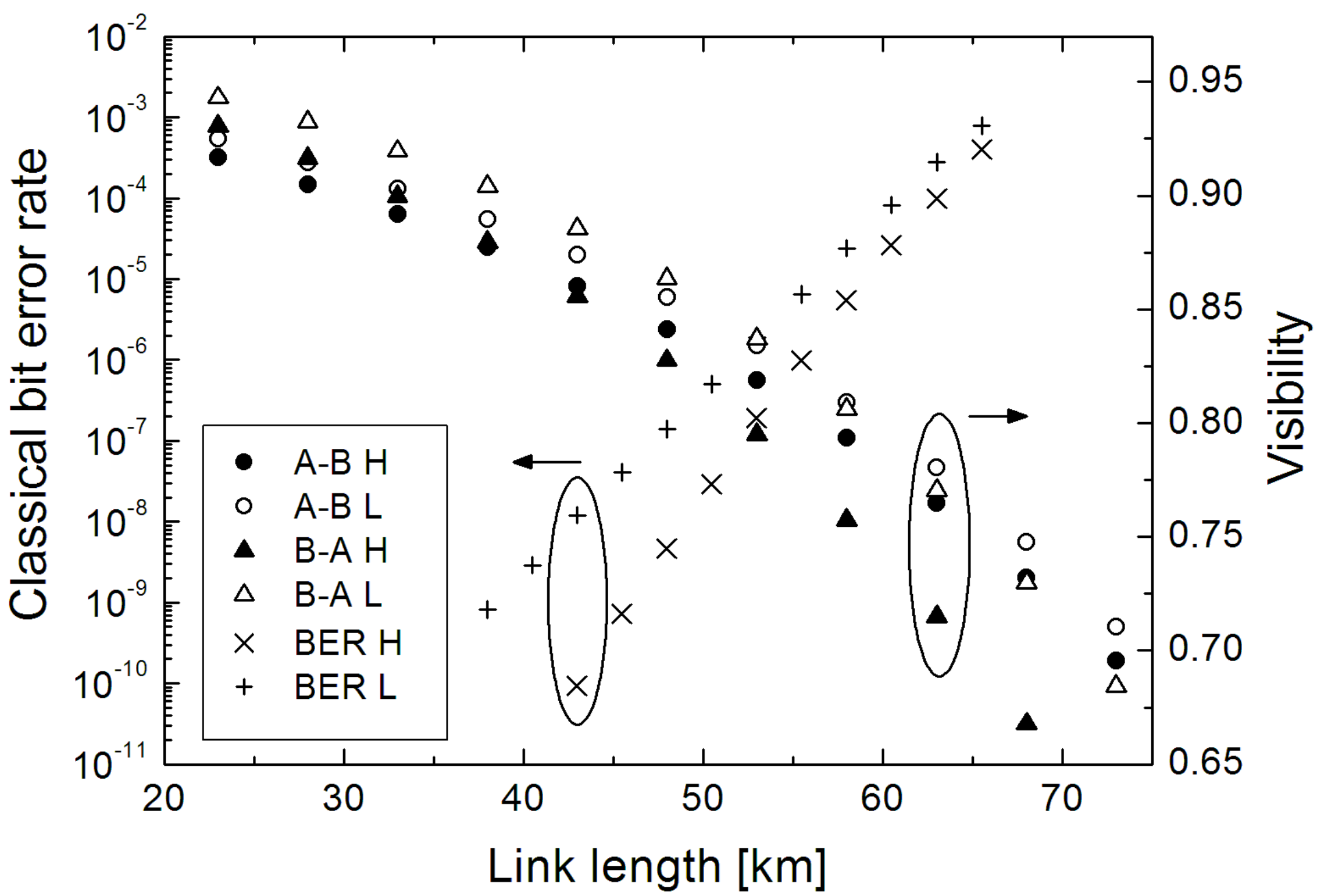